\documentclass[runningheads]{llncs}

\usepackage{amssymb}
\usepackage{fancyvrb}
\usepackage{bbm}
\usepackage[greek,english]{babel}
\usepackage{textgreek}
\usepackage{ucs}
\usepackage[utf8x]{inputenc}
\usepackage{autofe}
\usepackage{textcomp}
\usepackage{amsmath}

\DeclareUnicodeCharacter{8988}{\ensuremath{\ulcorner}}
\DeclareUnicodeCharacter{8989}{\ensuremath{\urcorner}}
\DeclareUnicodeCharacter{8803}{\ensuremath{\overline{\equiv}}}

\usepackage[demo]{graphicx}
\usepackage{caption}
\usepackage{subfigure}

\usepackage[belowskip=-10pt,aboveskip=-8pt]{caption}

\usepackage{url}
\urldef{\mailsa}\path||    
\newcommand{\keywords}[1]{\par\addvspace\baselineskip
\noindent\keywordname\enspace\ignorespaces#1}
\allowdisplaybreaks[4]
\newenvironment{ncenter}
 {\parskip=0pt\par\nopagebreak\centering}
 {\par\noindent\ignorespacesafterend}

\begin{document}

\mainmatter  

\title{Code Generation for Higher Inductive Types}
\subtitle{A Study in Agda Metaprogramming}
\titlerunning{A Study in Agda Metaprogramming}

\author{Paventhan Vivekanandan}
%
\authorrunning{P. Vivekanandan}
%
\institute{Indiana University, Bloomington, Indiana, USA\\
\email{pvivekan@umail.iu.edu}}

%
%

%


%
%

\toctitle{Lecture Notes in Computer Science}
\tocauthor{Authors' Instructions}
\maketitle

\begin{abstract}
Higher inductive types are inductive types that include nontrivial higher-dimensional structure, represented as identifications that are not reflexivity. While work proceeds on type theories with a computational interpretation of univalence and higher inductive types, it is convenient to encode these structures in more traditional type theories with mature implementations. However, these encodings involve a great deal of error-prone additional syntax. We present a library that uses Agda's metaprogramming facilities to automate this process, allowing higher inductive types to be specified with minimal additional syntax.
\end{abstract}

\section{Introduction}
\label{sec:intro}

Type theory unites programming and mathematics in a delightful synthesis, in which we can write programs and proofs in the same language.
Work on higher-dimensional type theory has revealed a beautiful higher-dimensional structure, lurking just beyond reach.
In particular, higher inductive types provide a natural encoding of many otherwise-difficult mathematical concepts, and univalence lets us work in our type theory the way we do on paper: up to isomorphism.
Homotopy type theory, however, is not yet done.
We do not yet have a mature theory or a mature implementation.

While work proceeds on prototype implementations of higher-dimensional type theories \cite{cohen}\cite{Angiuli-2017}, much work remains before they will be as convenient for experimentation with new ideas as Coq, Agda, or Idris is today.
In the meantime, it is useful to be able to experiment with ideas from higher-dimensional type theory in our existing systems.
If one is willing to put up with some boilerplate code, it is possible to encode higher inductive types and univalence using postulated identities. 

Boilerplate postulates, however, are not just inconvenient, they are also an opportunity to make mistakes.
Luckily, this boilerplate code can be mechanically generated using Agda's recent support for \emph{elaborator reflection}~\cite{David-2016}, a paradigm for metaprogramming in an implementation of type theory.
An elaborator is the part of the implementation that translates a convenient language designed for humans into a much simpler, more explicit, verbose language designed to be easy for a machine to process.
Elaborator reflection directly exposes the primitive components of the elaborator to metaprograms written in the language being elaborated, allowing them to put these components to new uses.

Homotopy type theory has thus far primarily been applied to the encoding of mathematics, rather than to programming.
Nevertheless, there are a few applications of homotopy type theory to programming. Applications such as homotopical patch theory \cite{Angiuli-2014} discuss a model of the core of the of Darcs \cite{Darcs-2005} version control system using patch theory \cite{Mimram-2013} encoded as a HIT. Containers in homotopy type theory~\cite{Altenkirch-2014,Abbott-2005} implement data structures such as multisets and cycles. Automating the HIT boilerplate code allows more programmers to begin experimenting with programming with HITs.

Using Agda's elaborator reflection, we automatically generate the support code for many useful higher inductive types, specifically those that include additional paths between constructors, but not paths between paths, which is sufficient for treating various interesting examples on the programming side \cite{Angiuli-2014}\cite{Paventhan-2018}\cite{Basold-2017}. We automate the production of the recursion principles, induction principles, and their computational behavior.
Angiuli et al.'s encoding of patch theory as a higher inductive type~\cite{Angiuli-2014} requires approximately 1500 lines of code when represented using rewrite mechanism. Using our library, the encoding can be expressed in just 70 lines.

This paper makes the following contributions:
\begin{itemize}
\item We describe the design and implementation of a metaprogram that automates an encoding of higher inductive types with one path dimension using Agda's new metaprogramming system.
\item We demonstrate applications of this metaprogram to examples from the literature, including both standard textbook examples of higher inductive types as well as larger systems, including both patch theory and specifying cryptographic schemes.
\item This metaprogram serves as an example of the additional power available in Agda's elaborator reflection relative to earlier metaprogramming APIs.
\end{itemize}

In Agda, we don't have built-in primitives to support the definition of higher inductive types. In this paper, we use Agda's rewrite rules mechanism to define higher inductive types \cite{Cockx16}\cite{Brunerie16}. Unlike \cite{Brunerie16}, we use basic modules, without parameters, to encode higher inductive types. This is because Agda's reflection library does not have primitives to support introducing parameterized modules.

\section{Background}
\label{sec:sec2}

\subsection{Higher Inductive Types}
\label{sec:sec2.1}

Homotopy type theory~\cite{HoTT-2013} is a research program that aims to develop univalent, higher-dimensional type theories.
A universe is \emph{univalent} when equivalences between types are considered equivalent to identifications between types. A type theory is univalent when every universe in the type theory is univalent; it is \emph{higher-dimensional} when we allow non-trivial identifications that every structure in the theory must nevertheless respect.
Identifications between elements of a type are considered to be at the lowest dimension, while identifications between identifications at dimension $n$ are at dimension $n+1$.
Voevodsky added univalence to type theories as an axiom, asserting new identifications without providing a means to compute with them.
While more recent work arranges the computational mechanisms of the type theory such that univalence can be derived, as is done in cubical type theories~\cite{cohen}\cite{Angiuli-2017}, we are concerned with modeling concepts from homotopy type theory in existing, mature implementations of type theory, so we follow Univalent Foundations Program \cite{HoTT-2013} in modeling paths using Martin-Löf's identity type.
Higher-dimensional structure can arise from univalence, but it can also be introduced by defining new type formers that introduce not only introduction and elimination principles, but also new non-trivial identifications.

In homotopy type theories, one tends to think of types not as collections of distinct elements, but rather through the metaphor of topological spaces.
The individual elements of the type correspond with points in the topological space, and identifications correspond to paths in this space.

While work proceeds on the general schematic characterization of higher inductive types\cite{Evan-2018}\cite{Basold-2017}\cite{Dybjer-2017}, it is convenient to syntactically represent the higher inductive types that we know are acceptable using a syntax similar to a traditional inductive type by providing its constructors (\textit{i.e.} its points); we additionally specify the higher-dimensional structure by providing additional constructors for paths. For example, Figure~\ref{fig:circle} describes \Verb|Circle|, which is a higher inductive type with one point constructor \Verb|base| and one non-trivial path constructor \Verb|loop|.
\begin{figure}[h]
\begin{center}
\begin{BVerbatim}
data Circle : Set where
  base : Circle
  loop : base ≡ base
\end{BVerbatim}
\end{center}
\caption{A specification of a higher inductive type}
\label{fig:circle}
\end{figure}
\normalsize


\begin{figure}
\begin{center}
\begingroup
\fontsize{8pt}{2pt}\selectfont
\begin{Verbatim}
postulate
  _↦_ : ∀ {i} {A : Set i} → A → A → Set i


{-# BUILTIN REWRITE _↦_ #-}



module Circle where



  postulate
    S : Set
    base : S
    loop : base ≡ base



  postulate
    recS : S → (C : Set) → (cbase : C) → (cloop : cbase ≡ cbase) → C
    βbase : (C : Set) → (cbase : C) → (cloop : cbase ≡ cbase) → 
      recS base C cbase cloop ↦ cbase


  {-# REWRITE βbase #-}



  postulate
    βloop : (C : Set) → (cbase : C) → (cloop : cbase ≡ cbase) → 
      ap (λ x → recS x C cbase cloop) loop ≡ cloop



  postulate
    indS : (x : S) → (C : S → Set) → 
      (cbase : C base) → (cloop : transport C loop cbase ≡ cbase) → C x
    iβbase : (C : S → Set) → 
      (cbase : C base) → (cloop : transport C loop cbase ≡ cbase) → 
      indS base C cbase cloop ↦ cbase


  {-# REWRITE iβbase #-}



  postulate
    iβloop : (C : S → Set) → 
      (cbase : C base) → (cloop : transport C loop cbase ≡ cbase) → 
      apd (λ x → indS x C cbase cloop) loop ≡ cloop
\end{Verbatim}
\endgroup
\end{center}
\caption{A HIT encoded using rewrite rules}
\label{fig:circle-hit}
\end{figure}
\normalsize

Figure \ref{fig:circle-hit} represents the implementation of {\tt Circle} in Agda. Inside module {\tt Circle}, the type {\tt S} and the constructors {\tt base} and {\tt loop} and the recursion and induction principles are declared as postulates.
{\tt recS} ignores the path argument and simply computes to the appropriate answer for the point constructor. 
The computation rule for point {\tt base} is declared as a rewrite rule using \Verb|{-# REWRITE , ...#-}| pragma. The computation rule for the path constructor {\tt loop} is postulated using reduction rule {\tt βloop}.
The operator \texttt{ap} is frequently referred to as \texttt{cong}, because it expresses that propositional equality is a congruence.
However, when viewed through a homotopy type theory lens, it is often called \texttt{ap}, as it describes the action of a function on paths.
In a higher inductive type, \texttt{ap} should compute new paths from old ones.
\begin{center}
\fontsize{8pt}{0pt}\selectfont
\begin{BVerbatim}
ap : {A B : Set} {x y : A} (f : A → B) (p : x ≡ y) → f x ≡ f y
\end{BVerbatim}
\end{center}

In addition to describing the constructors of the points and paths of \texttt{S}, Figure~\ref{fig:circle-hit} additionally demonstrates the dependent eliminator (that is, the induction rule) {\tt indS} and its computational meaning.
The dependent eliminator relies on another operation on identifications, called \texttt{transport}, that coerces an inhabitant of a family of types at a particular index into an inhabitant at another index.
Outside of homotopy type theory, \texttt{transport} is typically called \texttt{subst} or \texttt{replace}, because it also expresses that substituting equal elements for equal elements is acceptable.
\begin{center}
\fontsize{8pt}{0pt}\selectfont
\begin{BVerbatim}
transport : {A : Set} {x y : A} → (P : A → Set) → (p : x ≡ y) → P x → P y
\end{BVerbatim}
\end{center}

In the postulated computation rule for \texttt{indS}, the function {\tt apd} is the dependent version of \texttt{ap}: it expresses the action of dependent functions on paths.
\begin{center}
\fontsize{8pt}{0pt}\selectfont
\begin{BVerbatim}
apd : {A : Set} {B : A → Set} {x y : A} → (f : (a : A) → B a) → 
      (p : x ≡ y) → transport B p (f x) ≡ f y
\end{BVerbatim}
\end{center}

\subsection{Agda Reflection}
\label{sec:sec2.2}
Agda~\cite{norell:thesis} is a functional programming language with full dependent types and dependent pattern matching.
Agda's type theory has gained a number of new features over the years, among them the ability to restrict pattern matching to that subset that does not imply Streicher's Axiom K~\cite{Cockx2014}, which is inconsistent with univalence.
The convenience of programming in Agda, combined with the ability to avoid axiom K, makes it a good laboratory for experimenting with the idioms and techniques of univalent programming while more practical implementations of univalent type theories are under development.

Agda's reflection library enables compile-time metaprogramming.
This reflection library directly exposes parts of the implementation of Agda's type checker and elaborator for use by metaprograms, in a manner that is similar to Idris's elaborator reflection~\cite{davidphd,David-2016} and Lean's tactic metaprogramming~\cite{Ebner2017}.
The type checker's implementation is exposed as effects in a monad called \texttt{TC}.

Agda exposes a representation of its syntax to metaprograms, including datatypes for expressions (called \texttt{Term}) and definitions (called \texttt{Definition}).
The primitives exposed in \texttt{TC} include declaring new metavariables, unifying two \texttt{Term}s, declaring new definitions, adding new postulates, computing the normal form or weak head normal form of a \texttt{Term}, inspecting the current context, and constructing fresh names.
This section describes the primitives that are used in our code generation library; more information on the reflection library can be found in the Agda documentation \cite{Agda-doc-2017}.

\texttt{TC} computations can be invoked in three ways: by macros, which work in expression positions, using the \texttt{unquoteDecl} operator in a declaration position, which can bring new names into scope, and using the \texttt{unquoteDef} operator in a declaration position, which can automate constructions using names that are already in scope.
This preserves the principle in Agda's design that the system never invents a name.

An Agda \emph{macro} is a function of type {\tt $t_1$ → $t_2$ → $\ldots$ → Term → TC ⊤} that is defined inside a \texttt{macro} block.
Macros are special: their last argument is automatically supplied by the type checker and consists of a \texttt{Term} that represents the metavariable to be solved by the macro.
If the remaining arguments are quoted names or \texttt{Term}s, then the type checker will automatically quote the arguments at the macro's use site.
At some point, the macro is expected to unify the provided metavariable with some other term, thus solving it.

\begin{figure}[t]
\centering
\begin{minipage}{.5\textwidth}
\centering
\begingroup
\fontsize{8pt}{2pt}\selectfont
\begin{Verbatim}
macro
  mc1 : Term → Term → TC ⊤
  mc1 exp hole =
    do exp' ← quoteTC exp
       unify hole exp'



sampleTerm : Term
sampleTerm = mc1 (λ (n : Nat) → n)
\end{Verbatim}
\endgroup
\captionof{figure}{A macro that quotes its argument}
\label{fig:macro1}
\end{minipage}%
\begin{minipage}{.5\textwidth}
\centering
\begingroup
\fontsize{8pt}{2pt}\selectfont
\centering
\begin{Verbatim}
macro
  mc2 : Term → Term → TC ⊤
  mc2 exp hole =
    do exp' ← unquoteTC exp
       unify hole exp'



sampleSyntax : Nat → Nat
sampleSyntax =
  mc2 (lam visible (abs "n" (var 0 [])))
\end{Verbatim}
\endgroup
\caption{A macro that unquotes its argument}
\label{fig:macro2}
\end{minipage}
\end{figure}

Figure~\ref{fig:macro1} demonstrates a macro that quotes its argument.
The first step is to quote the quoted expression argument again, using \texttt{quoteTC}, yielding a quotation of a quotation.
This double-quoted expression is passed, using Agda's new support for Haskell-style do-notation, into a function that unifies it with the hole.
Because unification removes one layer of quotation, \texttt{unify} inserts the original quoted term into the hole.
The value of {\tt sampleTerm} is
\begin{center}
\begin{BVerbatim}
lam visible (abs "n" (var 0 []))
\end{BVerbatim}
\end{center}
The constructor \texttt{lam} represents a lambda, and its body is formed by the abstraction constructor \texttt{abs} that represents a scope in which a new name \texttt{"n"} is bound.
The body of the abstraction is a reference back to the abstracted name using de~Bruijn index \texttt{0}.

The {\tt unquoteTC} primitive removes one level of quotation.
Figure~\ref{fig:macro2} demonstrates the use of \texttt{unquoteTC}.
The macro {\tt mc2} expects a quotation of a quotation and substitutes its unquotation for the current metavariable.

The {\tt unquoteDecl} and {\tt unquoteDef} primitives, which run \texttt{TC} computations in a declaration context, will typically introduce new declarations by side effect.
A function of a given type is declared using {\tt declareDef}, and it can be given a definition using {\tt defineFun}.
Similarly, a postulate of a given type is defined using {\tt declarePostulate}.
Figure~\ref{fig:plus} shows an Agda implementation of addition on natural numbers, while Figure~\ref{fig:plusmeta} demonstrates an equivalent metaprogram that adds the same definition to the context.

\begin{figure}[t]
\centering
\begingroup
\fontsize{8pt}{2pt}\selectfont
\begin{Verbatim}
plus : Nat → Nat → Nat
plus zero b = b
plus (suc n) b = suc (plus n b)
\end{Verbatim}
\endgroup
\captionof{figure}{Addition on natural numbers}
\label{fig:plus}
\end{figure}
\begin{figure}
\centering
\begingroup
\fontsize{8pt}{2pt}\selectfont
\begin{Verbatim}
pattern vArg x = arg (arg-info visible relevant) x
pattern _`⇒_ a b = pi (vArg a) (abs "_" b)
pattern `Nat = def (quote Nat) []



unquoteDecl plus =
  do declareDef (vArg plus) (`Nat `⇒ `Nat `⇒ `Nat)
     defineFun plus
       (clause (vArg (con (quote zero) []) ::
                vArg (var "y") ::
                [])
          (var 0 []) ::
        clause (vArg (con (quote suc)
                       (vArg (var "x") :: [])) ::
                vArg (var "y") ::
                [])
          (con (quote suc)
            (vArg (def plus
                    (vArg (var 1 []) ::
                     vArg (var 0 []) :: [])) ::
             [])) ::
        [])
\end{Verbatim}
\endgroup
\caption{Addition, defined by metaprogramming}
\label{fig:plusmeta}
\end{figure}

In Figure~\ref{fig:plusmeta}, {\tt declareDef} declares the type of {\tt plus}.
The constructor {\tt pi} represents dependent function types, but a pattern synonym is used to make it shorter.
Similarly, \texttt{def} constructs references to defined names, and the pattern synonym \Verb|`Nat| abbreviates references to the defined name \texttt{Nat}, and {\tt vArg} represents the desired visibility and relevance settings of the arguments.
Once declared, \texttt{plus} is defined using {\tt defineFun}, which takes a name and a list of clauses, defining the function by side effect.
Each clause consists of a pattern and a right-hand side.
Patterns have their own datatype, while right-hand sides are \texttt{Term}s.
The name \texttt{con} is overloaded: in patterns, it denotes a pattern that matches a particular constructor, while in \texttt{Term}s, it denotes a reference to a constructor.

The next section introduces the necessary automation features by describing the automatic generation of eliminators for a variant on Dybjer's inductive families. Section 4 then generalizes this feature to automate the production of eliminators for higher inductive types using the rewrite mechanism. Section 5 revisits Angiuli et al.'s encoding of Darcs's patch theory \cite{Angiuli-2014} and demonstrates that the higher inductive types employed in that paper can be generated succinctly using our library\footnote{Please see \url{https://github.com/pavenvivek/WFLP-18}}.

\section{Code Generation for Inductive Types}
\label{sec:sec3}
An inductive type $D$ is a type that is freely generated by a finite collection of constructors. The constructors of $D$ accept zero or more arguments and have $D$ as the co-domain. The constructors can also take an element of type $D$ itself as an argument, but only \emph{strictly positively}: any occurrences of the type constructor $D$ in the type of an argument to a constructor of $D$ must not be to the left of any arrows.
Type constructors can have a number of \emph{parameters}, which may not vary between the constructors, as well as \emph{indices}, which may vary.

In Agda, constructors are given a function type. In Agda's reflection library, the constructor {\tt data-type} of the datatype {\tt Definition} stores the constructors of an inductive type as a list of \texttt{Name}s. The type of a constructor can be retrieved by giving its {\tt Name} as an input to the {\tt getType} primitive. In this section, we discuss how to use the list of constructors and their types to generate code for the elimination rules of an inductive type.

\subsection{Non-dependent Eliminators}
\label{sec:sec3.1}

\newcommand*{\KW}[1]{\textsf{\textbf{#1}}}
\newcommand*{\SET}{\textsf{Set}}

In Agda, we define an inductive type using {\tt data} keyword. A definition of an inductive datatype declares its type and specifies its constructors.
While Agda supports a variety of ways to define new data types, we will restrict our attention to the subset that corresponds closely to Dyber's inductive families. In general, the definition of an inductive datatype $D$ with constructors $c_1 \ldots c_n$ has the following form:
\begin{ncenter}
\fontsize{8pt}{2pt}\selectfont
\begin{align*}
&\KW{data}\ D\ (a_1 : A_1) \ldots (a_n:A_n) : (i_1 : I_1) \rightarrow\ldots\rightarrow (i_m : I_m) \rightarrow \SET{}\ \KW{where}  \\
&\hspace{0.3cm} c_1 : \Delta_1 \rightarrow D \, a_1 \ldots a_n \, e_{11} \ldots e_{1m}  \\
&\hspace{3.5cm} \vdots  \\
&\hspace{0.3cm} c_r : \Delta_n \rightarrow D \, a_1 \ldots a_n \, e_{r1} \ldots e_{rm}
\end{align*}
\end{ncenter}
where the index instantiations $e_{k1} \ldots e_{km}$ are expressions in the scope induced by the telescope $\Delta_k$. Every expression in the definition must also be well-typed according to the provided declarations. A telescope $\Delta \, = \, (x_1 : B_1) \ldots (x_n : B_n)$ is a sequence of types where later types may depend on elements of previous types. 


\begin{figure}
\begin{BVerbatim}
data Vec (A : Set) : Nat → Set where
  []   : Vec A zero
  _::_ : {n : Nat} → (x : A) → 
         (xs : Vec A n) → Vec A (suc n)

\end{BVerbatim}
\caption{Length-indexed lists}
\label{fig:vec}
\end{figure}

As an example, the datatype \texttt{Vec} (Figure~\ref{fig:vec}) represents lists of a known length. There is one parameter, namely \texttt{(A : Set)}, and one index, namely \texttt{Nat}. The second constructor, \texttt{\_::\_}, has a recursive instance of \texttt{Vec} as an argument.

While inductive datatypes are essentially characterized by their constructors, it must also be possible to eliminate their inhabitants, exposing the information in the constructors.
This section describes an Agda metaprogram that generates a non-dependent recursion principle for an inductive type; section~\ref{sec:sec3.2} generalizes this technique to fully dependent induction principles.

For {\tt Vec}, the recursion principle says that, in order to eliminate a \texttt{Vec A n}, one must provide a result for the empty \texttt{Vec} and a means for transforming the head and tail of a non-empty \texttt{Vec} combined with the result of recursion onto a tail into the desired answer for the entire \texttt{Vec}.
Concretely, the type of the recursor \texttt{recVec} is given as follows.

\begin{center}
\fontsize{8pt}{8pt}\selectfont
\begin{BVerbatim}
recVec : (A : Set) → {n : Nat} → Vec A n →
         (C : Set) → (base : C) →
         (step : {n : Nat} → (x : A) → (xs : Vec A n) → C → C) → C
\end{BVerbatim}
\end{center}

The recursor \texttt{recVec} maps the constructor {\tt []}, which takes zero arguments, to \texttt{base}.
It maps \texttt{(x :: xs)} to {\tt (step x xs (recVec xs C base step))}.
Because \texttt{step} is applied to a recursive call to the recursor, it takes one more argument than the constructor \texttt{\_::\_}.

Based on the schematic presentation of inductive types $D$ earlier in this section, we can define a schematic representation for their non-dependent eliminators $D_{\mathit{rec}}$.
\begin{ncenter}
\fontsize{8pt}{2pt}\selectfont
\begin{align*}
D_{\mathit{rec}} :\ & (a_1 : A_1) \to \ldots \to (a_n : A_n) \to\\
& (i_1 : I_1) \to \ldots \to (i_m : I_m) \to\\
& (\mathit{tgt} : D\ a_1 \ldots a_n\ i_1\ \ldots\ i_n) \to\\
& (\mathit{C} : \SET{}) \to \\
& (\mathit{f_1} : \Delta_1^\prime \to \mathit{C}) \to \ldots \to (\mathit{f_r} : \Delta_r^\prime \to \mathit{C}) \to \\
& \mathit{C}
\end{align*}
\end{ncenter}
The type of $f_i$, which is the method for fulfilling the desired type $C$ when eliminating the constructor $c_i$, is determined by the type of $c_i$.
The telescope $\Delta_i^\prime$ is the same as $\Delta_i$ for non-recursive constructor arguments.
However, $\Delta_i^\prime$ binds additional variables when there are recursive occurrences of $D$ in the arguments. For instance, if $\Delta_i$ has an argument $(y : B)$, where $B$ is not an application of $D$ or a function returning such an application, $\Delta_i^\prime$ binds $(y : B)$ directly.
If $B$ is an application of $D$, then an additional binding $(y^\prime : C)$ is inserted following $y$. Finally, if $B$ is a function type $\Psi \to D$, the additional binding is $(y^\prime : \Psi \to C)$. 

To construct the type of \texttt{recVec}, we need to build the types of {\tt base} and {\tt step}.
These are derived from the corresponding types of \texttt{[]} and \texttt{\_::\_}, which can be discovered using reflection primitives.
Since {\tt []} requires no arguments, its corresponding method is {\tt (base : C)}.
The constructor {\tt pi} of type {\tt Term} encodes the abstract syntax tree (AST) representation of {\tt \_::\_}. We can retrieve and traverse the AST of {\tt \_::\_}, and add new type information into it to build a new type representing {\tt step}. Once the AST for {\tt step}'s type has been found, it is possible to build the type of {\tt recVec}. To quantify over the return type \texttt{(C : Set)}, we use the \texttt{Term} constructor \texttt{agda-sort} to refer to \texttt{Set}. 


In general, when automating the production of $D_{\mathit{rec}}$, all the information that is needed to produce the type signature is available in the \texttt{TC} monad by looking up $D$'s definition.
The constructor {\tt data-type} contains the number of parameters occurring in a defined type. It also encodes the constructors of the type as a list of \texttt{Name}s. Metaprograms can retrieve the index count by using the type and the number of parameters. The constructors of $D$ refer to the parameter and the index using de~Bruijn indices.

The general schema for the computation rules corresponding to $D_{\mathit{rec}}$ and constructors $c_1, \ldots, c_n$ is as follows:
\newcommand*{\RHSap}[2]{\ensuremath{\mathsf{RHS}\left(#1, #2\right)}}
\newcommand*{\RHS}[1]{\RHSap{f_{#1}}{\Delta_{#1}^\prime}}
\begin{ncenter}
\fontsize{8pt}{2pt}\selectfont
\begin{align*}
& D_\mathit{rec}\ a_1\ \ldots\ a_n\ i_1\ \ldots\ i_m\ (c_1\ {\Delta_1}) \, C \, f_1 \ldots f_r = \RHS{1}\\
& \vdots  \\
& D_\mathit{rec}\ a_1\ \ldots\ a_n\ i_1\ \ldots\ i_m\ (c_r\ {\Delta_r}) \, C \, f_1 \ldots f_r = \RHS{r}
\end{align*}
\end{ncenter}
\normalsize

Here, ${\Delta_j}^\prime$ is the sequence of variables bound in $\Delta_j$.
$\mathsf{RHS}$ constructs the application of the method $f_j$ to the arguments of $c_j$, such that $C$ is satisfied.
It is defined by recursion on $\Delta_j$.
$\RHSap{f_j}{\cdot}$ is $f_j$, because all arguments have been accounted for.
$\RHSap{f_j}{(y : B)\Delta_k}$ is $\RHSap{f_j\ y}{\Delta_k}$ when $B$ does not mention $D$.
$\RHSap{f_j}{(y : D) (y^\prime : C)\Delta_k}$ is $\RHSap{f_j\ y\ \left(D_{\mathit{rec}}\ \ldots y \ldots\right)}{\Delta_k}$, where the recursive use of $D_{\mathit{rec}}$ is applied to the recursive constructor argument as well as the appropriate indices, and the parameters, result type, and methods remain constant.
Higher-order recursive arguments are a generalization of first-order arguments.
Finally, $\RHSap{f_j}{(y : \Psi \to D) (y^\prime : \Psi \to C)\Delta_k}$ is $\RHSap{f_j\ y\ \left(\lambda \overline{\Psi} . D_{\mathit{rec}}\ \ldots \left(y\ \overline{\Psi} \right) \ldots\right)}{\Delta_k}$ where the recursive use of $D_{\mathit{rec}}$ is as before.


After declaring \texttt{recVec}'s type using {\tt declareDef}, it is time to define its computational meaning using the schematic rules defined above.
The computation rule representing the action of function \texttt{recVec} on {\tt []} and {\tt \_::\_} is defined using {\tt clause}. The first argument to {\tt clause} encodes variables corresponding to the above type, and it also includes the abstract representation of the constructors {\tt []} and {\tt \_::\_} on which the pattern matching should occur. The second argument to \texttt{clause}, which is of type {\tt Term}, refers to the variables in the first argument using de~Bruijn indices, and it encodes the output of \texttt{recVec} when the pattern matches.
The computation rules for \texttt{recVec} are given as follows.

\begin{center}
\fontsize{8pt}{8pt}\selectfont
\begin{Verbatim}
    recVec []        C base step = base
    recVec (x :: xs) C base step = step x xs (f xs C base step)
\end{Verbatim}
\end{center}

{\tt generateRec} (Figure~\ref{fig:LibM}) build the computation and elimination rules respectively. The recursion rule generated by {\tt generateRec} is brought into scope using {\tt unquoteDecl}. The first argument to {\tt generateRec} is the quoted {\tt Name} of the recursor encoded inside {\tt Arg}, and the second argument is the quoted {\tt Name} of the inductive type.


\begin{figure}[t]
\centering
\fontsize{8pt}{2pt}\selectfont
\begin{Verbatim}
generateRec, generateInd : Arg Name → (indType : Name) → TC ⊤



generateβRec, generateβInd : Arg Name → List (Arg Name) →
  (indType : Name) → (param : Nat) → (points : List Name) → TC ⊤



generateRecHit, generateIndHit : Arg Name → 
  (indType : Name) → (baseElim : Name) → (param : Nat) → 
  (points : List Name) → (paths : List Name) → TC ⊤





generateβRecHitPath, generateβIndHitPath : Name → List (Arg Name) →   
  (indType : Name) → (baseElim : Name) → (param : Nat) → 
  (points : List Name) → (paths : List Name) → TC ⊤


\end{Verbatim}
\captionof{figure}{Library for generating dependent and non-dependent eliminators}
\label{fig:LibM}
\end{figure}

\subsection{Dependent Eliminators}
\label{sec:sec3.2}

The dependent eliminator for a datatype, also known as the \emph{induction principle}, is used to eliminate elements of a datatype when the type resulting from the elimination mentions the very element being eliminated. The type of the induction principle for $D$ is:
\begin{ncenter}
\fontsize{8pt}{2pt}\selectfont
\begin{align*}
D_{\mathit{ind}} :\ & (a_1 : A_1) \to \ldots \to (a_n : A_n) \to\\
& (i_1 : I_1) \to \ldots \to (i_m : I_m) \to\\
& (\mathit{tgt} : D\ a_1 \ldots a_n\ i_1\ \ldots\ i_m) \to\\
& \begin{array}{@{}l@{}l}(\mathit{C} :\ & (i_1 : I_1) \to \ldots \to (i_m : I_m) \to\\  & D\ a_1 \ldots a_n\ i_1\ \ldots\ i_n \to \SET{})\to\end{array} \\
& (\mathit{f_1} : \Delta_1^\prime \to \mathit{C}\ e_{11} \ldots e_{1p}\ (c_1\ \overline{\Delta_1})) \to \\
& (\mathit{f_r} : \Delta_r^\prime \to \mathit{C}\ e_{r1} \ldots e_{rp}\ (c_r\ \overline{\Delta_r})) \to \\
& \mathit{C}\ i_1\ \ldots\ i_n\ \mathit{tgt}
\end{align*}
\end{ncenter}
Unlike the non-dependent recursion principle $D_{\mathit{rec}}$, the result type is now computed from the target and its indices.
Because it expresses the reason that the target must be eliminated, the function $C$ is often referred to as the \emph{motive}.
Similarly to  $D_{\mathit{rec}}$, the type of each method $f_i$ is derived from the type of the constructor $c_i$---the method argument telescope $\Delta_k^\prime$ is similar, except the arguments that represents the result of recursion now apply the motive $C$ to appropriate arguments.
If $\Delta_i$ has an argument $(y : B)$, where $B$ is not an application of $D$ or a function returning such an application, $\Delta_i^\prime$ still binds $(y : B)$ directly.
If $B$ is an application of $D$ to parameters $a\ldots$ and indices $e\ldots$, then an additional binding $(y^\prime : C\ e\ldots\ y)$ is inserted following $y$.
Finally, if $B$ is a function type $\Psi \to D\ a\ldots\ e\ldots$, the additional binding is $(y^\prime : \Psi \to C\ e\ldots (y\ \overline{\Psi}))$. 

Following these rules, the induction principle for \texttt{Vec} can be defined as follows.

\begin{ncenter}
\fontsize{8pt}{8pt}\selectfont
\begin{BVerbatim}


indVec : (A : Set) → {n : Nat} → (xs : Vec A n) → 
         (C : {n : Nat} → Vec A n → Set) → (base : C []) → 
         (step : {n : Nat} → (x : A) → 
                 (xs : Vec A n) → C xs → C (x :: xs)) → C xs

\end{BVerbatim}
\end{ncenter}

Automating the production of the dependent eliminator is an extension of the procedure for automating the production of the non-dependent eliminator.
The computation rules for the induction principle are automated using the same approach as for the recursion principle.
The generation of induction principles is carried out using {\tt generateInd} (Figure~\ref{fig:LibM}).

%

\section{Code Generation for Higher Inductive Types}
\label{sec:sec4}

In Agda, there are no built-in primitives to support the definition of higher inductive types.
However, we can still define a higher inductive type using rewrite rules, as described in section~\ref{sec:sec2.1}.
In this section, we discuss the automation of code generation for the elimination and the computation rules of higher inductive types. While the general formulation of higher inductive types is a subject of active research \cite{Dybjer-2017}\cite{Lumsdaine-2017}\cite{Ambrus-2018}, we stick to a schema that follows the pattern of Basold et al.'s \cite{Basold-2017} general rules for higher inductive types.

\subsection{Non-dependent Eliminators for HITs}
\label{sec:sec4.1}

The recursion principle of a higher inductive type $G$ maps the points and paths of $G$ to points and paths in an output type $C$. We extend the general schema of the recursion principle given in section~\ref{sec:sec3.1} by adding methods for path constructors (Figure~\ref{fig:Grec}).

\begin{figure}[t]
\centering
\begin{minipage}{.5\textwidth}
\centering
\fontsize{8pt}{2pt}\selectfont
\begin{align*}
G_{\mathit{rec}} :\ & (a_1 : A_1) \to \ldots \to (a_n : A_n) \to\\
& (i_1 : I_1) \to \ldots \to (i_m : I_m) \to\\
& (\mathit{tgt} : G\ a_1 \ldots a_n\ i_1\ \ldots\ i_n) \to\\
& (\mathit{C} : \SET{}) \to \\
& (\mathit{f_1} : \Delta_1^\prime \to \mathit{C}) \ldots (\mathit{f_r} : \Delta_r^\prime \to \mathit{C}) \to \\
& (\mathit{k_1} : \Delta_1^\prime \to (f_i \ldots) \equiv (f_j \ldots)) \to \\
& \vdots \\
& (\mathit{k_q} : \Delta_q^\prime \to (f_i \ldots) \equiv (f_j \ldots)) \to \\
& \mathit{C}
\end{align*}
\captionof{figure}{Generic schema for recursor}
\label{fig:Grec}
\end{minipage}%
\begin{minipage}{.5\textwidth}
\centering
\fontsize{8pt}{2pt}\selectfont
\begin{align*}
\beta G_{rec} :\ & (a_1 : A_1) \to \ldots \to (a_n : A_n) \to\\
& (\mathit{C} : \SET{}) \to \\
& (\mathit{f_1} : \Delta_1^\prime \to \mathit{C}) \ldots (\mathit{f_r} : \Delta_r^\prime \to \mathit{C}) \to \\
& (\mathit{k_1} : \Delta_1^\prime \to (f_i \ldots) \equiv (f_j \ldots)) \to \\
& \vdots \\
& (\mathit{k_q} : \Delta_q^\prime \to (f_i \ldots) \equiv (f_j \ldots)) \to \\
& \mathit{ap} \  (\lambda \  x . G_{rec} \  x \  C \  f_1 \  \ldots\ f_r \  k_1 \ \ldots\  k_q)\\
& \hspace{0.25cm} \  (p_i \ldots) \equiv (k_i \ldots)
\end{align*}
\captionof{figure}{Generic schema for computation rule corresponding to Grec}
\label{fig:BGrec}
\end{minipage}
\end{figure}


The schematic definition of $G_{\mathit{rec}}$ supports only one-dimensional paths.
The type of the method $f_i$ for a point constructor $g_i$ in $G_{rec}$ is built the same way as for the normal inductive type $D$, as described in section~\ref{sec:sec3.1}.
The code generator builds the type of $k_i$, method for path constructor $p_i$ in $G_{rec}$, by traversing the AST of $p_i$. The arguments of $k_i$ are handled the same way as for the point constructor's method $f_i$. During the traversal, the code generator uses the base type recursor $D_{rec}$ to map the point constructors $g_i$ of $G$ in the codomain of $p_i$ to $f_i$. Determining the computation rules corresponding to points $g_i$ is similar to the computation rules corresponding to constructors $c_i$ of the inductive type $D$, except that there are additional methods to handle paths. Paths compute new paths; the computation rules that govern the interaction of recursors and paths $p_i$ are named and postulated. They identify the action of the recursor on the path with the corresponding method. The computation rules corresponding to paths $p_i$ are postulated as given in Figure~\ref{fig:BGrec}.

As an example, if the code for the circle HIT from section~\ref{sec:sec2.1} has been generated, and the type is called \texttt{S}, then the recursor needs a method for \texttt{base} and one for \texttt{loop}.
The method for \texttt{base} should be an inhabitant of \texttt{C}.
If it is called \texttt{cbase}, then the method for \texttt{loop} should be a path \texttt{cbase ≡ cbase}.
The types of the path methods depend on the values of the point methods.
The code generator builds the type of \texttt{loop}'s method by traversing the AST of \texttt{loop}'s type,
replacing references to point constructors with the result of applying the base type's recursor to the point methods. The recursion rule {\tt recS} follows this pattern.

\begin{center}
\fontsize{8pt}{2pt}\selectfont
\begin{BVerbatim}
  recS : S → (C : Set) → (cbase : C) → (cloop : cbase ≡ cbase) → C
\end{BVerbatim}
\end{center}
\normalsize

The code generator builds the computation rule for the point constructor {\tt base} using the same approach as described in section~\ref{sec:sec3.1} as if it were for the base type.
Additionally, it includes variables in the {\tt clause} definition for the path constructor {\tt loop}.
The code generator postulates the following computation rule {\tt βloop} for the path constructor {\tt loop}:

\begin{center}
\fontsize{8pt}{2pt}\selectfont
\begin{BVerbatim}
    βloop : (C : Set) → (cbase : C) → (cloop : cbase ≡ cbase) → 
      ap (λ x → recS x C cbase cloop) loop ≡ cloop
\end{BVerbatim}
\end{center}
\normalsize

The application of function {\tt recS} to the path {\tt loop} substitutes the point {\tt base} for the argument {\tt x} and it evaluates to the path {\tt cloop} in the output type {\tt C}. In the tool, {\tt generateRecHit} is used to build the elimination rule and the computation rules for points, and {\tt generateβRecHitPath} is used to build the computation rules for paths (Figure~\ref{fig:LibM}). The third argument to {\tt generateRecHit} is the base type's recursor built using {\tt generateβRec} that constructs the computation rules for points using rewrite rules. The parameter count is passed as the fourth argument.

\subsection{Dependent Eliminators for HITs}
\label{sec:sec4.2}

The dependent eliminator for a higher inductive type $G$ is a dependent function that maps an element $g$ of $G$ to an output type $C \, g$. The general schema for the induction principle of $G$ is given in Figure~\ref{fig:Gind}.

\begin{figure}[t]
\centering
\begin{minipage}{.5\textwidth}
\centering
\fontsize{8pt}{2pt}\selectfont
\begin{align*}
G_{\mathit{ind}} :\ & (a_1 : A_1) \to \ldots \to (a_n : A_n) \to\\
& (i_1 : I_1) \to \ldots \to (i_m : I_m) \to\\
& (\mathit{tgt} : G\ a_1 \ldots a_n\ i_1\ \ldots\ i_n) \to\\
& \begin{array}{@{}l@{}l}(\mathit{C} :\ & (i_1 : I_1) \to \ldots \to (i_m : I_m) \to\\  & G\ a_1 \ldots a_n\ i_1\ \ldots\ i_n \to\\ &\SET{})\to\end{array} \\
& (\mathit{f_1} : \Delta_1^\prime \to \mathit{C}\ j_{11} \ldots j_{1p}\ (c_1\ \overline{\Delta_1})) \to \\
& \vdots\\
& (\mathit{f_r} : \Delta_r^\prime \to \mathit{C}\ j_{r1} \ldots j_{rp}\ (c_r\ \overline{\Delta_r})) \to \\
& (\mathit{k_1} : \Delta_1^\prime \to \texttt{transport} \, \mathit{C} \, p_1 \, (f_i \ldots) \\
&\hspace{1.5cm} \equiv (f_j \ldots)) \to \\
& \vdots\\
& (\mathit{k_q} : \Delta_q^\prime \to \texttt{transport} \, \mathit{C} \, p_q \, (f_i \ldots) \\
&\hspace{1.5cm} \equiv (f_j \ldots)) \to \\
& \mathit{C}\ i_1\ \ldots\ i_n\ \mathit{tgt}
\end{align*}
\captionof{figure}{Generic schema for induction principle}
\label{fig:Gind}
\end{minipage}%
\begin{minipage}{.5\textwidth}
\centering
\fontsize{8pt}{2pt}\selectfont
\begin{align*}
\beta G_i :\ & (a_1 : A_1) \to \ldots \to (a_n : A_n) \to\\
& \begin{array}{@{}l@{}l}(\mathit{C} :\ & (i_1 : I_1) \to \ldots \to (i_m : I_m) \to\\  & G\ a_1 \ldots a_n\ i_1\ \ldots\ i_n \to\\ &\SET{})\to\end{array} \\
& (\mathit{f_1} : \Delta_1^\prime \to \mathit{C}\ j_{11} \ldots j_{1p}\ (c_1\ \overline{\Delta_1})) \to \\
& (\mathit{f_r} : \Delta_r^\prime \to \mathit{C}\ j_{r1} \ldots j_{rp}\ (c_r\ \overline{\Delta_r})) \to \\
& (\mathit{k_1} : \Delta_1^\prime \to \texttt{transport} \, \mathit{C} \, p_1 \, (f_i \ldots) \\
&\hspace{1.5cm} \equiv (f_j \ldots)) \to \\
& (\mathit{k_r} : \Delta_r^\prime \to \texttt{transport} \, \mathit{C} \, p_r \, (f_i \ldots) \\
&\hspace{1.5cm} \equiv (f_j \ldots)) \to \\
& \texttt{apd} \  (\lambda \  x\ .\ G_{ind} \  x \  C \  f_1 \ \ldots\ f_r \  k_1 \ \ldots\ k_r)\\
&\hspace{0.5cm} \  (p_i \ldots) \equiv\\
& (k_i \ldots)
\end{align*}
\captionof{figure}{Generic schema for computation rule corresponding to Gind}
\label{fig:BGind}
\end{minipage}
\end{figure}

Similar to $G_{rec}$, the type of $f_i$ is built the same way as for the normal inductive type $D$.
The code generator builds the type of the method for path constructor $p_i$, called $k_i$, in $G_{ind}$, by traversing the AST of $p_i$. During the traversal, the code generator uses the base eliminator $D_{ind}$ to map the point constructors $g_i$ of $G$ in the codomain of $p_i$ to $f_i$. In the first argument to the identity type in the codomain of $k_i$, the code generator adds an application of  {\tt transport} to the motive {\tt C} and the path $p_i$. The arguments of $k_i$ are handled the same way as for $f_i$.
The computation rules corresponding to paths $p_i$ are postulated as given in Figure~\ref{fig:BGind}.

For the type {\tt S} with point constructor {\tt base} and path constructor {\tt loop}, to define a mapping {\tt indS : (x : S) → C x}, we need {\tt cbase : C base} and {\tt cloop : transport C loop cbase ≡ cbase}, where {\tt cloop} is a heterogeneous path transported over {\tt loop}. The code generator builds the type of {\tt cloop} by adding relevant type information to the type of {\tt loop}.
The type of the method for the path constructor {\tt cloop} is derived by inserting a call to {\tt transport} with arguments {\tt C}, {\tt loop}, and {\tt cbase}. 
The code generator applies the base eliminator to map the point {\tt base} to {\tt cbase} during the construction of the codomain of {\tt cloop}. The following declaration gives the type of {\tt indS}.
\begin{center}
\fontsize{8pt}{2pt}\selectfont
\begin{BVerbatim}
indS : (circle : S) → (C : S → Set) → (cbase : C base) →
  (cloop : transport C loop cbase ≡ cbase) → C circle
\end{BVerbatim}
\end{center}

The computation rule for {\tt base}, which defines the action of {\tt indS} on {\tt base}, is built using the same approach as for the non-dependent eliminator {\tt recS}.
The postulated computation rule {\tt iβloop} for the path {\tt loop} uses {\tt apd} which gives the action of dependent function {\tt indS} on the path {\tt loop}.
\begin{center}
\fontsize{8pt}{2pt}\selectfont
\begin{BVerbatim}
iβloop : (C : S → Set) → (cbase : C base) → 
  (cloop : transport C loop cbase ≡ cbase) → 
  apd (λ x → indS x C cbase cloop) loop ≡ cloop
\end{BVerbatim}
\end{center}

{\tt generateIndHit} is used to build the elimination rule and the computation rules for points, and {\tt generateβIndHitPath} is used to build the computation rules for paths (Figure~\ref{fig:LibM}). 

\section{Applications}

\subsection{Patch Theory Revisited}
\label{patch-theory}

We reimplemented Angiuli et al.'s patch theory \cite{Angiuli-2014} using our code generator in Agda. We implemented basic patches such as the insertion of a string as line $l_1$ in a file and deletion of a line $l_2$ from a file. The functions implementing insertion and deletion in the universe are not bijective. So, we used Angiuli et al.'s patch history approach to encode non-bijective functions. According to this approach, we developed a separate higher inductive type \texttt{History} which serves as the types of patches. We also implemented patches involving encryption or decryption with cryptosystems like RSA and Paillier. In addition to easing the implementation difficulties of higher inductive types, the code generator greatly reduced the code size. The type definitions shrank from around 1500 to around 70 lines, resulting in a 60\% decrease in the overall number of lines of code in the development.

\subsection{Cryptographic Protocols}
\label{crypto}

Vivekanandan \cite{Paventhan-2018} models certain cryptographic protocols using homotopy type theory,  introducing a new approach to formally specifying cryptographic schemes using types. The work discusses modeling cryptDB~\cite{Popa-2011} using a framework similar to Angiuli et al.'s patch theory. CryptDB employs layered encryption techniques and homomorphic encryption. We can implement cryptDB by modeling the database queries as paths in a higher inductive type and mapping the paths to the universe using singleton types \cite{Angiuli-2014}. The code generator can be applied to generate code for the higher inductive type representing cryptDB and its corresponding elimination and computation rules. By using the code generator, we can decrease the length and increase the readability of the definitions, hopefully making it more accessible to the broad cryptographic community.

\section{Related Work}

Kokke and Swierstra \cite{Kokke-2015} implemented a library for proof search using Agda's old reflection primitives, from before it had elaborator reflection. They describe a Prolog interpreter in the style of Stutterheim et al. \cite{Stutterheim-2013}. It employs a hint database and a customizable depth-first traversal, with lemmas to assist in the proof search.

Van der Walt and Swierstra \cite{Walt-2013} and van der Walt \cite{Walt-2012} discuss automating specific categories of proofs using proof by reflection.
A key component of this proof technique is a means for converting an expression into a quoted representation.
They automate this process, giving a user-defined datatype. Van der Walt \cite{Walt-2012} also give an overview of Agda's old metaprogramming tools.

Datatype-generic programming \cite{Altenkirch-2006}\cite{Loh-2011}\cite{Chapman-2010} via universes allows defining a single function over an entire class of datatypes at once, saving developers the effort of implementing the operation for datatypes specific to their programs. As Agda's reflection library evolves and the internal representation of datatypes changes, the tool described in this paper requires maintenance work. A future direction would be to work with a universe extended with support for higher inductive types. In such case, the only metaprograms necessary are those that convert to and from the universe. The metaprograms automating the elimination rules do not need to change as long as the universe is kept the same.

Ongoing work on cubical type theories~\cite{cohen}\cite{Angiuli-2017}\cite{cubicalHIT} provides a computational interpretation of univalence and HITs.
We strenuously hope that these systems quickly reach maturity, rendering our code generator obsolete.
In the meantime, however, these systems are not yet as mature as Agda.

\section{Conclusion and Future Work}

We presented a code generator that generates the encodings of higher inductive types, developed using Agda's new support for Idris-style elaborator reflection.
In particular, the tool generates the dependent and non-dependent elimination rules and the computational rules for 1-dimensional higher inductive types.
This syntax is greatly simplified with respect to writing the encoding by hand.
We demonstrated an extensive reduction in code size by employing our tool. Next, we intend to extend the tool to support higher-dimensional paths in the definition of HITs, bringing its benefits to a wider class of problems.

\section*{Acknowledgements}

The author is greatly indebted to David Christiansen for his contributions and advice, and the anonymous reviewers for their valuable review comments.

\end{document}